\documentclass[aps, showpacs, onecolumn, superscriptaddress, preprint]{revtex4}

\usepackage{here}
\usepackage{color}
\usepackage{graphicx}
\usepackage{epstopdf}
\usepackage{amsmath}        
\usepackage{amssymb}        
\usepackage{mathrsfs}       
\usepackage[mathscr]{eucal}     
\usepackage{bm}             
\usepackage{verbatim}           
\usepackage{ulem}

\newcounter{myromancnt}

\begin{document}

\title{Plasma Dipole Oscillation Excited by Trapped Electrons Leading to Bursts of Coherent Radiation}
\author{Kyu Been Kwon}
\affiliation{School of Natural Science, UNIST, 50 UNIST-gil, Ulju-gun, Ulsan, 689-798, Korea}
\author{Teyoun Kang}
\affiliation{School of Natural Science, UNIST, 50 UNIST-gil, Ulju-gun, Ulsan, 689-798, Korea}
\author{Hyung Seon Song}
\affiliation{School of Natural Science, UNIST, 50 UNIST-gil, Ulju-gun, Ulsan, 689-798, Korea}
\author{Young-Kuk Kim}
\affiliation{School of Natural Science, UNIST, 50 UNIST-gil, Ulju-gun, Ulsan, 689-798, Korea}
\author{Bernhard Ersfeld}
\affiliation{Department of Physics, Scottish Universities Physics Alliance and University of Strathclyde, Glasgow G4 0NG, UK}
\author{Dino A. Jaroszynski}
\affiliation{Department of Physics, Scottish Universities Physics Alliance and University of Strathclyde, Glasgow G4 0NG, UK}
\author{Min Sup Hur}
\email{mshur@unist.ac.kr}
\affiliation{School of Natural Science, UNIST, 50 UNIST-gil, Ulju-gun, Ulsan, 689-798, Korea}

\date{\today}

\begin{abstract}
Plasma dipole oscillation (PDO) depicted as harmonic motion of a spatially localized block of electrons has,  until now, been  hypothetical. In practice, the plasma oscillation occurs always as a part of a plasma wave. Studies on radiation burst from plasmas  have focused only on coupling of the plasma wave and electromagnetic wave. 
Here we show that  a very-high-field PDO can be generated by the electrons trapped in a moving train of potential wells. The electrons riding on the  potential train coherently construct  a local dipole moment  by charge separation. The subsequent PDO is found to persist stably until its energy is emitted entirely via coherent radiation.
 In our novel method, the moving potentials are provided by 
two slightly-detuned laser pulses colliding in a non-magnetized plasma. 
The radiated energy reaches  several millijoules in the terahertz spectral region. The proposed method  provides a way of realizing the PDO  as a new radiation source in the laboratory. 
PDO as a mechanism of astrophysical radio-bursts is discussed.
\end{abstract}
\pacs{52.59.Ye, 52.25.Os, 52.65.Rr}

\maketitle


Plasma oscillations are usually described as in-phase motion of a spatially-localized block of electrons  (plasma dipole oscillations, PDO). However, so far the PDO has been hypothetical; there has been no efficient method to generate it, and its  stability  is  not known. In practice, plasma oscillations  always  occur as  parts of a travelling  plasma wave, which is a collection of infinitesimal plasma oscillators with different phases.   
 A plasma wave in a non-magnetized, homogeneous plasma cannot emit an electromagnetic wave, since, as  there is no crossing point of their  dispersion curves, the energy exchange between those waves is prevented. Hence, previous studies have focused on finding mechanisms for how the electrostatic plasma wave is  coupled to an electromagnetic wave in inhomogeneous or magnetized plasmas, to explain the observed radio-bursts  from the solar corona \cite{reid,boyd,sturrock,melrose,yoon},  the terahertz, or higher-frequency emission from laboratory plasmas \cite{chinfatt,hamster,yoshii,yugami,sheng,jjkim,gopal,cho,liao,liao2,bezzerides,teubner,quere,ijkim}.  Known  wave-wave coupling mechanisms include linear mode conversion and coherent wake emission \cite{sheng,quere}, parametric plasmon  coalescence \cite{melrose,teubner},  and Cherenkov wake emission \cite{yoshii,yugami}.   However, the PDO has not been explored  as a mechanism of  the radiation emission.  
\begin{figure}[b]  
\centering
\includegraphics[scale=0.6]{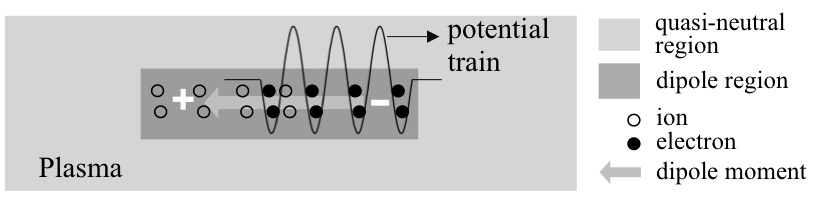}
\caption{Generation of a dipole moment in a plasma  by the electrons trapped in a moving potential train.}
\end{figure}

In this Letter, we show, for the first time, that a very high-field PDO  can be generated without relying on external magnetic or electric fields. Also we show that the PDO persists stably until all of its energy is emitted  via a burst  of coherent radiation, which is distinct  from the emission by conventional wave-wave coupling.  The  mechanism to generate  a PDO, revealed first here,  is  building up   a dipole moment by electrons trapped in a spatially localized, moving train of  potential wells. 
The trapped electrons, riding on the potential train,  are {\it displaced}  to generate a dipole field by charge separation between the displaced electrons and the remaining ions (Fig. 1).
Later, as the moving potentials diminish,  the dipole  is  {\it released}  to commence  plasma oscillation, which serves as the dipole radiator. 
As the electron trapping is prevalent in plasmas of diverse scales, the PDO may be relevant to various known observations of the radio-bursts  from space (e.g., the coronal type III bursts \cite{reid}), or terahertz bursts from laser-solid interactions \cite{liao,liao2}.

\begin{figure}[b]  
\centering
\includegraphics[scale=0.55]{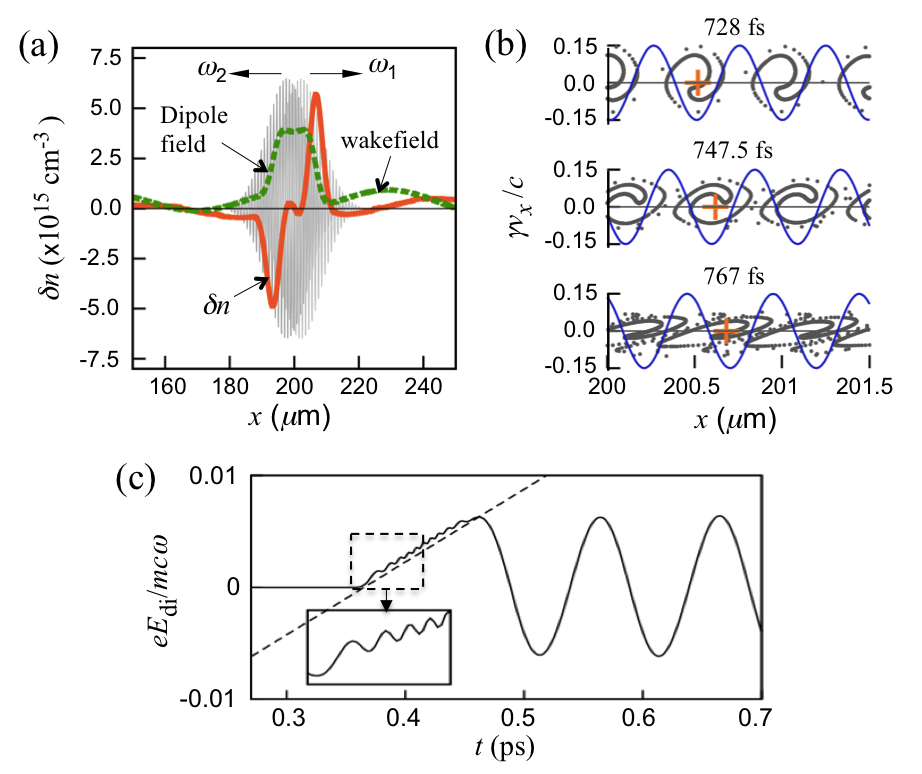}
\caption{(Color online) Displacement and release of trapped electrons from a 1D PIC simulation with normalized peak amplitude $a_0=eE_0/(mc\omega)=0.1$ ($\omega=[\omega_1+\omega_2]/2$)  and pulse duration $\tau=30$ fs for both laser pulses, unperturbed plasma density $n_0=3.125\times10^{17}~\rm cm^{-3}$, and wavelengths of the laser pulses, $\lambda_1=1~\mu\rm m$ and $\lambda_2=0.967~\mu\rm m$.  
(a) Perturbation of the electron density $\delta n$ (orange solid) by the colliding laser pulses, and the resulting electric field (green dashed).  (b) Trapped electrons and the PM potential (blue solid) at three different moments.   Crosses:  centre-of-mass of electrons in each potential well. The PM potential is given in arbitrary units.
(c) Time evolution of $E_{\rm di}$ spatially averaged over 4 $\mu\rm m$ around the centre of the pulse collision.  
}
\end{figure}

As a novel method to produce the moving potential train, here we consider the ponderomotive (PM) potential associated with the beat wave of two slightly-detuned  laser pulses that collide in the plasma.   In  the  region  where the pulses  overlap, electrons are trapped  by the PM potentials  that moves in the direction of the higher frequency pulse at a phase velocity $v_\phi=|\omega_1-\omega_2|/(k_1+k_2)$,  where $\omega_{1,2}$ and $k_{1,2}$ are the angular frequencies and wavenumbers of the laser pulses, respectively.  
The ``displacement-release'' process for this system  is numerically demonstrated using  one-dimensional (1D)  particle-in-cell (PIC) simulations (Fig. 2). 
The electron displacement yields a pair of oppositely-charged layers  (Fig. 2a, solid orange), which produces a dipole field (Fig. 2a, dashed green).  The distance between the  layers is comparable with the pulse  width. For obliquely colliding pulses, this distance is also controlled by the collision angle. The electron phase-space distributions close to the centre of the pulse collision (Fig. 2b), demonstrate  that the trapped electrons co-move with the PM potential.   As the trapped electrons move with constant $v_\phi$, the dipole electric field $E_{\rm di}(t)$  at the centre  increases linearly over time (Fig. 2c). The small, fast oscillation at the bounce frequency $\omega_b=2\omega a_0$  during the linear growth 
(Fig. 2c, inset) confirms  that the field growth is led by trapped electrons.   As will be shown, the emitted radiation from the PDO generated by colliding pulses exhibits a unique spectrum that includes  both a super-continuum and a narrow-bandwidth peak.  The energy of the radiation reaches  several millijoules in the terahertz spectral region,  making  it suitable for applications that require strong terahertz fields.

The dipole field can be generated in a similar way by colliding {\it chirped} pulses, which is easy to realize in practice \cite{vieux}.  In this case, in general, the growth of the dipole field is not linear with time.  Note that the large charge separation is a result of coherent displacement of the trapped electrons in multiple PM potential wells, which grow in numbers as the overlap of the laser pulses increases.   
On either side  of the dipole field, there is a low-amplitude wakefield  generated by the laser pulses before collision.  

The maximum displacement of electrons at the centre can be calculated  explicitly using a force-balance model (Fig. 3a).  
When  the normalized peak amplitude of the laser pulses, $a_0$ is sufficiently high, the PM force  increases faster, at an early stage, than the restoring force of the dipole field, thereby driving the electron displacement.  Later, when  the PM force drops below the restoring force, electron displacement stops and the electron dipole is released (Fig. 3a, point $R$).  When the time of the release is denoted by $t_{rel}$ (Fig. 3a), the maximum dipole  field is given by 
\begin{equation}
\label{Epk}
E_{\rm max}=\int_0^{t_{rel}}S(t)dt,
\end{equation}
where $S(t)=dE_{\rm di}(t)/dt$, i.e. the slope of the dipole field growth.  With an assumption that locally all the electrons are trapped, from Gauss's law, 
\begin{equation}
\label{slope}
S=\frac{en_0}{\epsilon_0}v_\phi=\frac{en_0}{\epsilon_0}\frac{\Delta\omega}{2k},
\end{equation}
where  $-e$ is the charge of a single electron, $\epsilon_0$ is the vacuum permittivity, $\Delta\omega=|\omega_1-\omega_2|$, and $2k=k_1+k_2$.    
This model predicts that, by increasing $a_0$ and the pulse duration $\tau$, $t_{rel}$ can be delayed considerably, leading to  very large $E_{\rm max}$, although the rate of increase of $E_{\rm max}$ becomes very slow for  relativistic laser intensities ($a_0\gg 1$). 

The growth of $E_{\rm di}(t)$ led by the trapped electrons should be preceded by wavebreaking of the slow plasma wave, which is driven by the initial weak portion of the PM force (Fig. 3a). The electron trapping follows the wavebreaking with almost no delay, since the plasma wave grows rapidly. To determine  $t_{rel}$, we first calculate $t_w$, which is defined by time from the start of wavebreaking to that of the maximum PM force.  Before wavebreaking, electrons  are driven into harmonic motion by the PM force, {\it i.e.}, $(\partial_t^2+\omega_p^2)x=-c^2ka^2$, where $\omega_p$ is the plasma frequency, $a$ the normalized amplitude of the laser pulses, and $x$ the electron displacement.  For longitudinally Gaussian laser pulses,  $a^2=\frac{1}{2}a_0^2\exp[-2(t-t_w)^2/\tau^2]e^{i\phi_{\rm PM}}+c.c.$, where $\phi_{\rm PM}=2kx-\Delta\omega t$.  Note that the wavebreaking occurs at $t=0$.  With $x=\frac{1}{2}\hat{x}e^{i\phi_{\rm PM}}+\rm c.c.$ (slow plasma wave), where $\hat{x}$ is the oscillation amplitude,  the equation of motion yields  
\begin{equation}
\label{vhat-eqmotion}
\frac{\partial^2\hat{x}}{\partial t^2}+\omega_p^2\hat{x}=-c\omega a_0^2\exp\left[-\frac{2(t-t_w)^2}{\tau^2}\right].
\end{equation}
When $|\hat{x}(t=0)|>1/(k_1+k_2)$,  wavebreaking occurs due to the change in ordering of electrons \cite{dawson}.   In Eq. (\ref{vhat-eqmotion}), we assume  $\hat{x}$ increases rapidly until  wavebreaking, so $\partial_t^2\hat{x}\gg \Delta\omega\partial_t\hat{x}\gg \Delta\omega^2\hat{x}$  (verified by PIC simulations), which is opposite to the regular assumption of the slowly-varying envelope models, {\it e.g.},  \cite{hurprl}.  $\hat{x}(t=0)$ can be calculated using a convolution integral of the Green's function and  source term of Eq. (\ref{vhat-eqmotion}) for $t=(-\infty,0)$. 
From the wavebreaking condition,
\begin{equation}
\label{tw}
\frac{8}{\omega^2\tau^2a_0^2}\simeq\frac{e^{-2\xi^2}}{\xi^2+\omega^2\tau^2/16}~,~~\xi=\frac{t_w}{\tau}.
\end{equation} 
\begin{figure}[b]  
\centering
\includegraphics[scale=0.62]{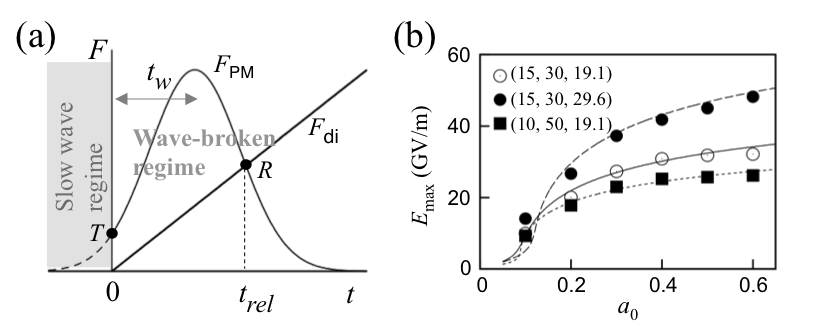}
\caption{(a) Evolution of the ponderomotive ($F_{\rm PM}$) and electric restoring forces ($F_{\rm di}$).  $T$ and $R$ are the wavebreaking threshold and release point of the dipole, respectively. 
(b) $E_{\rm max}$ vs. $a_0$ from simulations (points) and Eq. (\ref{balance}) (lines). The numbers in parentheses are $f_p$ ($\omega_p/2\pi$) (THz), the pulse duration $\tau$ (fs), and detuning $\Delta f$ ($=\Delta\omega/2\pi$) (THz).}
\end{figure}

The release time $t_{rel}$ can be obtained by balancing the ensemble-averaged PM force with the restoring force ($-eE_{\rm max}$). 
When the frequency difference of the laser pulses is constant, $E_{\rm max}=St_{rel}$, leading to 
\begin{equation}
\label{balance}
\frac{\omega_p^2}{2\omega^2}\Delta\omega t_{rel}=\eta\frac{a_0^2}{\sqrt{1+a_0^2/2}}\exp\left[-\frac{2(t_{rel}-t_w)^2}{\tau^2}\right],\\
\end{equation}
where $\eta$ is related to the ensemble average of the PM force exerted on electrons scattered  inside each PM potential segment. The ensemble average is equivalent to the time-average of the PM force exerted on a single electron during one bounce cycle  ({\it i.e.}, $x=A\sin\omega_bt+x_0$, where $A$ and $x_0$ are  the amplitude and centre of the bouncing motion, respectively),  leading to $\eta=J_0(2kA)\sin2kx_0$,
where $J_0$ is the zero'th order Bessel function. 
From  $A\le \lambda_b/2=\pi/2k$ and $\sin 2kx_0\le 1$,  $\eta\le 0.3$.   $E_{\rm max}$ from numerical solutions of Eqs. (\ref{Epk}), (\ref{slope}), (\ref{tw}), and (\ref{balance}) along with  $\eta=0.3$ agrees well with  1D PIC simulation data for a diverse range of parameters (Fig. 3b).  
Note that although we considered the electrons located initially at $x=0$ to obtain those equations, they are still valid for nearby electrons  for short-enough driving pulses. 
\begin{figure}[b]  
\centering
\includegraphics[scale=0.26]{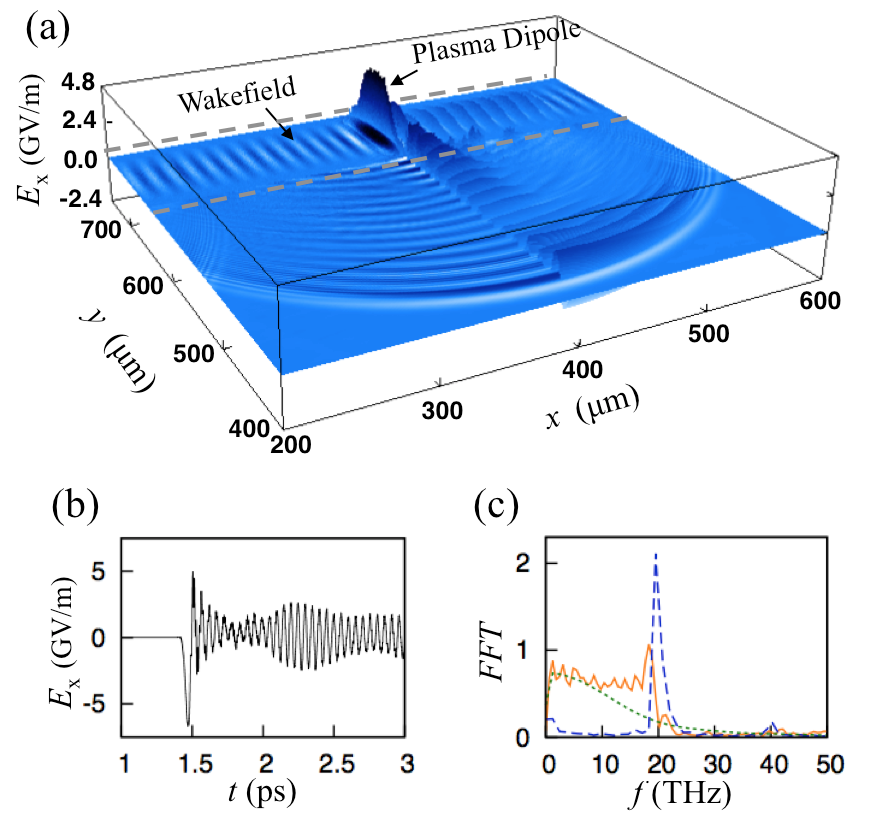}
\caption{(Color online) Two-dimensional PIC simulation of the radiation emission.  (a) The plasma is located in the region enclosed by the dashed lines. The plasma density is uniform along $x$, and trapezoidal in $y$ with $10~\mu\rm m$-ramps at both ends and a $100~\mu\rm m$-plateau. (a) The  left half ($x<400~\mu\rm m$) of $E_x$ (the radiation field) is presented by a contour plot for clear view of the radiation pattern.   (b) Evolution of $E_x$ measured at 60  $\mu\rm m$ from the dipole centre. (c) Fourier transform of (b) over $1.2\le t\le 1.5$ ps (green dotted), $1.2\le t\le 2$ ps (orange solid), and $2\le t\le 3.5$ ps (blue dashed). }
\end{figure}

To determine the radiation emission pattern from the PDO, we have performed two-dimensional (2D) PIC simulations.  We present the results for  $\omega_p/2\pi=20$ THz ($n_0=4.96\times10^{18}~\rm cm^{-3}$), $\lambda_1=0.8~\mu\rm m$, $\lambda_2=0.759~\mu\rm m$ ($\Delta f=20$ THz), $a_0=0.7$, $\tau=30$ fs, and spot radius 40 $\mu$m (Fig. 4a).  The laser pulses counter-propagate along the $x$-direction and  collide with  zero collision angle at the  centre of a plasma strip. The high amplitude dipole oscillation emits  bi-directional radiation (only one part is shown).  The peak amplitude of the radiation field reaches 5 GV/m determined at 60 $\mu$m from the centre of the dipole (Fig. 4b). When the laser pulses collide deep in a bulk plasma instead of  a strip, the duration of the burst  is prolonged with  the leading peak amplitude being lowered.  The radiation from the travelling wakefields observed on both sides of the dipole  is very weak because of destructive interference.   
The radiation field increases linearly over $1.4\le t\le 1.5$ ps during the pulse collision, and oscillates subsequently over $1.5\le t\le 2$  ps (Fig. 4b).    Interestingly, monochromatic radiation at $\omega_p$  is emitted even after the decay  of the dipole radiation (Fig. 4b, $t>2$ ps; 4c, dashed blue); this persisting emission is due to the collision of the travelling wakefields.   Appearance of this oscillation in the tail does not show consistent dependence on laser and plasma parameters:  it is sometimes prominent, as in the given case, but can also be negligible even under similar parameters.  
The spectral properties are noteworthy: the spectrum of the radiation from $t=1.2$ ps to the first decay ($\sim$ 2 ps) contains both a spectral continuum and a narrow-band peak at $\omega_p$ (Fig. 4c, solid orange). The low frequency part in the continuum (Fig. 4c, dotted green) results mostly from the first half cycle (Fig. 4b, $t=1.4\sim1.5$ ps), which corresponds to the linearly  growing phase  of the dipole field. This implies that the continuous spectrum can be controlled by frequency chirp of the driving pulses.  
\begin{figure}[b]  
\centering
\includegraphics[scale=0.65]{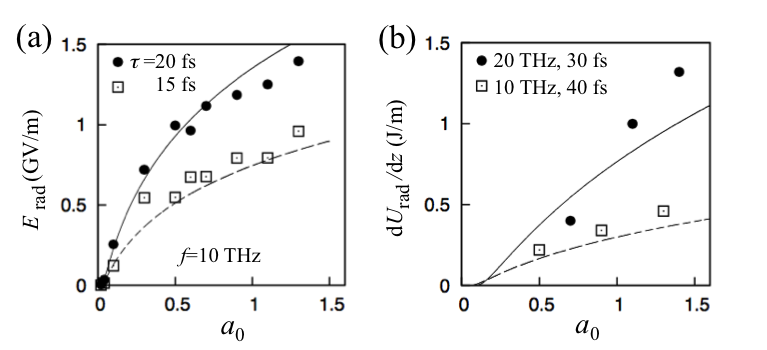}
\caption{Two dimensional simulations and theory of (a)  $E_{\rm rad}$ and (b) $dU_{rad}/dz$  vs.  $a_0$.  (a) $n_0=1.24\times 10^{18}~\rm cm^{-3}$ (10 THz), $\tau=$20 fs (circles), and 15 fs (boxes), $\lambda_1=0.8 ~\mu\rm m$, and  $\lambda_2=0.779 ~\mu\rm m$.  (b) The upper curve  (circles) is for $n_0=4.96\times10^{18}~\rm cm^{-3}$ (20 THz), $\tau=30$ fs, $\lambda_2=0.759~\mu\rm m$;  the lower curve  (squares) is for $1.24\times10^{18}~\rm cm^{-3}$, $\tau=40$ fs, $\lambda_2=0.75~\mu\rm m$. Throughout (a) and (b), the laser pulse spot is $40~\mu\rm m$.  In (b), the energy is integrated until the radiation field drops to $1/e^2$ of the leading peak. }
\end{figure}

The radiation amplitude $E_{\rm rad}$  in 2D space can be deduced from the 1D model of the dipole field.  In 2D polar coordinates, the electric field of radiation from 2D dipole moment is represented by  
\begin{equation*}
E_{\rm rad}=\mu_0\sqrt{\frac{c\omega^3}{8\pi}}\frac{\cos\phi}{\sqrt{r}}\left|\iint\vec{P}dxdy\right|,
\end{equation*}
where $r$ is radial distance, $\phi$ is polar angle, $\omega$ is the radiation frequency ($=\omega_p$),  and $\vec{P}$ is the electric polarization.   
As the charge is separated approximately with $2v_\phi$ over $t_{rel}$, $\int dx \simeq 2v_\phi t_{rel}=\frac{2\epsilon_0}{en_0}E_{\rm max}(y)$, where $E_{\rm max}$ is determined by Eq. (\ref{Epk}) at each $y$.  In 2D, the dipole field and polarization are related by  $\vec{P}+2\epsilon_0\vec{E}=0$.  Then we obtain  
\begin{equation}
\label{Erad2Dfinal}
E_{\rm rad}=\sqrt{\frac{2e^2}{\pi m^2c^3\omega_p}}\frac{\cos\phi}{\sqrt{r}}\int_{-\infty}^\infty E_{\rm  max}^2(y) dy.
\end{equation}
Equation (\ref{Erad2Dfinal}) agrees  well with 2D simulation data for  $\phi=0$ and $r=60~\mu\rm m$ (Fig. 5a).  In these simulations,  the spot radius of  the driver pulses is set to 40 $\mu\rm m$ and the plasma strip is  80 $\mu\rm m$ wide.

To estimate  the  radiation energy, we assume all the initial dipole field energy  $U_{\rm di}$ is converted to radiation energy $U_{\rm rad}$. Then, in 2D, 
\begin{equation}
\label{Udipole}
\frac{dU_{\rm rad}}{dz}~[{\rm J/m}]\simeq\frac{dU_{\rm di}}{dz}=\frac{\epsilon_0^2}{2en_0}\int_{-\infty}^\infty E_{\rm max}^3dy.
\end{equation}
Predictions of Eq. (\ref{Udipole}) agree  well with results of 2D PIC  simulations at $f_p=10$ and 20 THz (Fig. 5b).  The reasonable agreement implies that virtually all of the initial dipole energy is converted into radiation. 
Total radiation energy in 3D  can be approximated by multiplying  the lateral dimension ($z$-length) of the dipole to Eq. (\ref{Udipole}). 
To maximize the emission energy, the lateral dimension should be made as large as possible  by using highly elliptical driving pulses.  High-power elliptical laser  beams with an  eccentricity of  more than an order of magnitude are readily available \cite{shim,laskin}.  Assuming a $40~\mu{\rm m}\times 1000~\mu\rm m$ laser spot, the total energy of the dipole radiation is expected to reach  a few millijoules (Fig. 5b). This energy can be increased further by increasing the driving pulse intensity. 
 Also the system can be optimized  by adjusting the frequency chirping or the collision angle of the laser pulses.
 
PDO can be relevant to emission from diversely different plasma systems.  As the ejection of energetic electron beams (e.g., by magnetic reconnection \cite{egedal})  is prevalent in astrophysical plasmas (e.g., in the solar corona \cite{reid}), pairs of colliding beams  could  possibly generate dipole moments by trapping electrons in  plasma waves, which  grow from stream instabilities.   Similar dipole radiation in the terahertz spectral region may have been produced (but overlooked)  in laser-solid interactions \cite{liao,liao2}, where the reflected pulse overlaps  the incident pulse obliquely in the underdense plasma expanded from the surface of the solid.  In this system, the Doppler-shift caused by the moving reflection surface of the plasma can provide the reflected and incident pulses with the necessary frequency-detuning. The PDO driven by colliding pulses can also be used for a diagnostics of  local plasma density; as the radiation spectrum of the PDO  peaks at the plasma frequency,  the radiation contains the information of the local plasma density at the position of the pulse collision. 

The  mechanism of dipole generation  here is distinct from Ref. \cite{cho}, where  a weak dipole field is generated by a small nonlinear current that is {\it disrupted} by trapped electrons. 
 In contrast, here  the trapped electrons {\it generate} a   very strong dipole field  beyond wavebreaking; previous limitations in growth of the dipole field are circumvented.    Moreover, unlike Ref. \cite{cho},  the radiation can be directly emitted from the dipole without relying on an external magnetic field.  The mechanism we propose is also different from  coalescence of two plasmons \cite{melrose,teubner}, which generate second harmonic radiation. In our method, the  fundamental harmonic radiation is dominant. An oscillating plasma surface \cite{bulanov,lichters,norreys,dromey2,dromey3} is also a 
kind of  localized plasma oscillation, but it is limited  to a relatively steep plasma-vacuum interface. Moreover, differently from the PDO, it acts as a reflector  for  high harmonic generation,  rather than as a  radiation emitter.  Particle trapping by counter-propagating pulses was studied in the long-pulse (both or one of the them) regime \cite{hurprl,sheng2,shvets,shvets2}, but a localized PDO and the radiation from that  were not perceived.

In conclusion,  we have investigated the generation of localized plasma dipole oscillations (PDO) in non-magnetized plasmas by trapped electrons in a train of moving potential wells. As an efficient way to provide the necessary potential train, we considered the ponderomotive potentials generated by the beat of colliding laser pulses. 
It was shown that the PDO can persist stably until its entire energy is emitted via coherent radiation burst. 
Results of 2D PIC simulations suggest that the radiation energy can reach several millijoules  in the terahertz spectral region, which may lead to applications in terahertz sciences where very strong fields are required.   
The force-balance model predicts that the radiation energy can be increased considerably  by increasing the driving pulse intensity. Also, the pulse duration, frequency chirping, and collision angle of the pulses can be  further optimized. The suggested method can be used for diagnostics of local plasma density.  
The stably-persisting PDO is relevant to a radiation mechanism in diverse scales of plasma and over a large wavelength range,   
such as the  laser-solid interactions and radiation bursts from astrophysical plasmas.

\begin{acknowledgments}
This work was supported by the Basic Science Research Program through the National Research Foundation (NRF) of Korea funded by the Ministry of Science, ICT and Future Planning (Grant number NRF-2014M1A7A1A01030175 and NRF-2016R1A5A1013277). For the simulations, we were supported by the PLSI supercomputing resources of the Korea Institute of Science and Technology Information. DAJ and BE were  supported by the UK Engineering and Physical Sciences Council Grants no. EP/J018171/1 and EP/N028694/1, the ECs LASERLAB-EUROPE (grant agreement no. 284464, Seventh Framework Programme), EuCARD-2 (grant no. 312453, FP7) and the Extreme Light Infrastructure (ELI) European Project.  We used {\it VisIt} for visulisation of 2D data (https://wci.llnl.gov/simulation/computer-codes/visit).  We acknowledge Prof. D. Ryu at UNIST and Prof. Y.U. Jeong at KAERI for useful discussions and comments. 
\end{acknowledgments}

\end{document}